\documentclass[11pt,twoside]{article}

\usepackage{asp2006}
\usepackage{epsf}
\usepackage{psfig}
\usepackage{lscape}

\begin{document}
\title{The Metallicity Dependence of Wolf-Rayet Mass Loss}
\author{Andrew Onifer, Alexander Heger, and
Joseph Abdallah}
\affil{Los Alamos National Laboratory, PO Box 1663, Los Alamos, NM 87545, USA}

\begin{abstract}
We produce models of early WN, WC, and WO stars as a function
of metallicity $Z$ using an analytic CAK-type approach. 
At log~$Z / Z_\odot
\geq -2$ both WN and WC stars have the approximate dependence $\dot{M} \propto Z^{0.5}$.  For a WN wind the
mass-loss rate drops rapidly below log~$Z /
Z_\odot = -2$, and no wind solution can be found for log~$Z / Z_\odot < -3$.
For WC and WO winds the mass-loss rate plummets in the range $-3 \leq \mathrm{log}~Z
/ Z_\odot \leq -2$ and
tends to flatten due to the self-enrichment of C and O to around $10^{-8} \: M_\odot$ yr$^{-1}$ for log~$Z / Z_\odot
\leq -4$. No significant difference in $\dot{M}$ was found for
WC versus WO stars at low metallicity.
\end{abstract}

\section{Introduction}
Near solar metallicity the mass-loss rates of Wolf-Rayet (WR) stars can be strongly dependent on
$Z$, but the self-enrichment of CNO elements creates an
effective $Z$ floor that can allow large $\dot{M}$ even in a low-$Z$ environment.
Using the modified CAK approach described in
\citet{onifergayley06}, we calculate $\dot{M}$ as a function of $Z$ for
early-type WR winds from WN, WC, and WO stars.  To ensure a complete line list at
the high critical-point temperatures of these winds, we use data from the Kurucz
list \citep{kurucz79}, the Opacity Project \citep{seaton05}, and the opacity
group at Los Alamos \citep{mazevetabdallah06}.  The results are compared to the
$\dot{M}$-$Z$ dependences for late-type stars by \citet[hereafter
VdK05]{vinkdekoter05}.

\section{Models}
At the wind critical point we set the wind speed $v_c = 500$ km s$^{-1}$, the temperature $T_c
= 1.3\times10^5$ K, and the electron density $n_e = 10^{13}$ cm$^{-3}$.  The
luminosity is set such that the Eddington parameter $\Gamma = 0.5$.  The WN
model mass $M = 19 \: M_\odot$ and the terminal speed $v_\infty = 1700$ km
s$^{-1}$, corresponding to the WN4 star WR6 \citep{hamannetal06}.  For the WC
model $M = 14.5 \: M_\odot$ and $v_\infty = 2500$ km s$^{-1}$, corresponding to
the WC4 star HD 32125 \citep{crowtheretal02}. The
important abundances are shown by mass in Table \ref{table:abund}.  For the WN
model $X_\mathrm{N}$ was kept fixed, and for the WC and WO models $X_\mathrm{CNO}$ and
$X_\mathrm{Ne}$ were fixed at
all $Z$. The WO model uses the ratios C/He and O/He from the Sand 1 model in
\citet{kingsburghetal95} and is otherwise the same as the WC model. We account
for the effects of ionization stratification
\citep{lucyabbott93} by using a radiation thermalization length $L_{th}$ that
spans the temperature range $6.0\times10^4 K \leq T \leq 1.3\times10^5 K$ and
comparing $\dot{M}$ to the case where $L_{th} = 0$.

\begin{table}[!ht]
\caption{Wind mass fractions at $Z_\odot$\label{table:abund}}
\smallskip
\begin{center}
{\small
\begin{tabular}{cccccc}
\tableline
\noalign{\smallskip}
WR Type & $X_{\mathrm{He}}$ & $X_\mathrm{C}$ & $X_\mathrm{N}$ & $X_\mathrm{O}$ &
$X_{\mathrm{Ne}}$\\
\noalign{\smallskip}
\tableline
\noalign{\smallskip}
WN & 0.99 & $2.9\times10^{-4}$ &
$1.0\times10^{-2}$ & $1.8\times10^{-4}$ & $1.1\times10^{-3}$\\
\noalign{\smallskip}
WC & 0.48 & 0.43 & 0.0 & $6.4\times10^{-2}$ & $2.1\times10^{-2}$\\
WO & 0.21 & 0.51 & 0.0 & 0.25 & $2.1\times10^{-2}$\\
\noalign{\smallskip}
\tableline
\end{tabular}
}
\end{center}
\end{table}

\section{Results and Conclusions}
\begin{figure}[!ht]
\plottwo{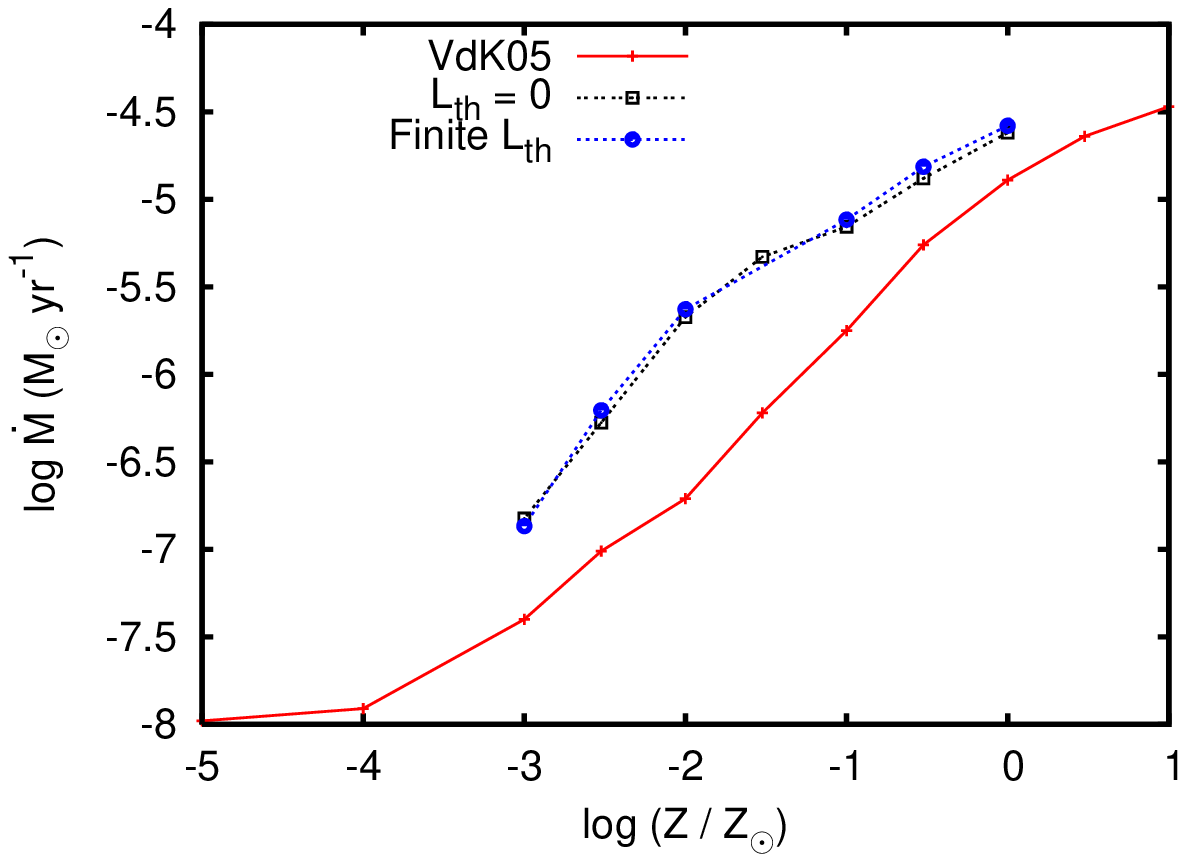}{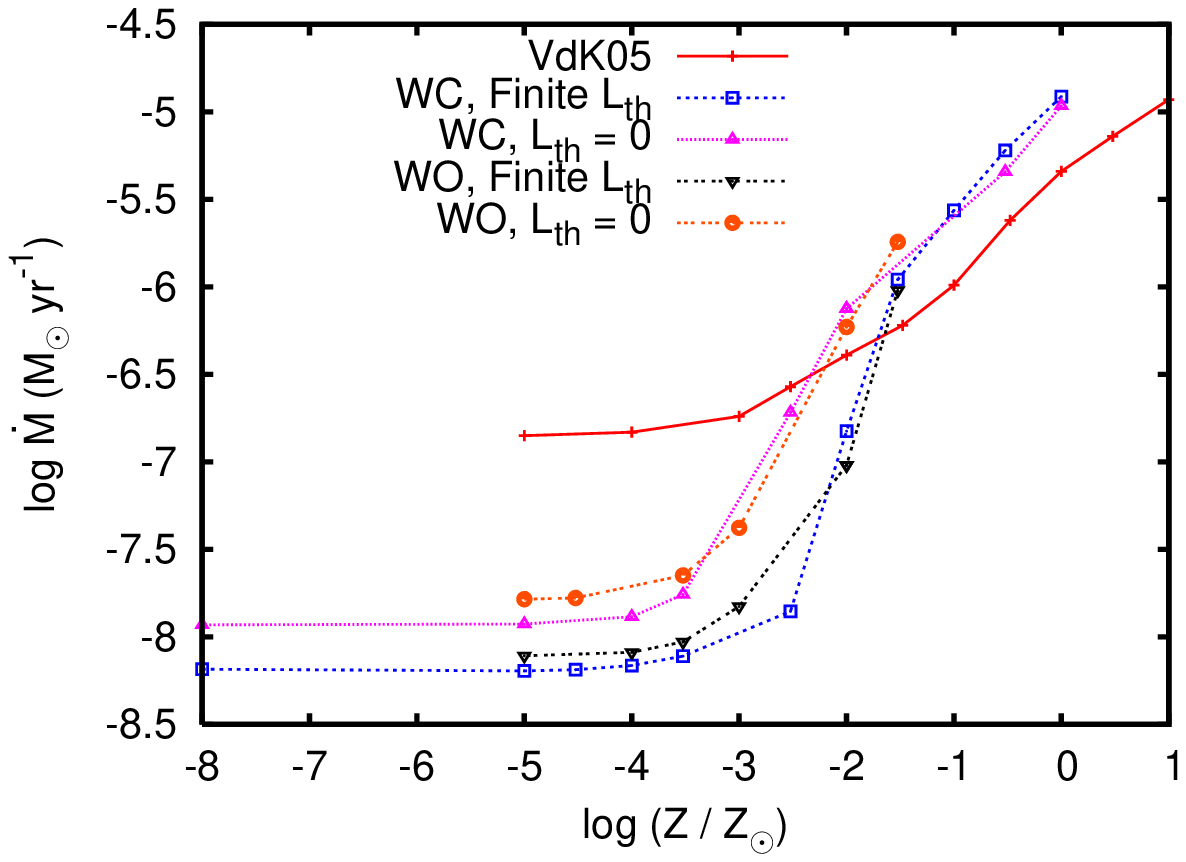}
\caption{Left: WN $\dot{M}$ vs $Z$ for finite $L_{th}$ and $L_{th}$ = 0
Right: WC and WO $\dot{M}$ with finite $L_{th}$
and $L_{th}$ = 0.  The VdK05 results (solid lines) are also shown.
\label{fig:wnwc}}
\end{figure}

Figure \ref{fig:wnwc} shows the $\dot{M} \propto Z^m$ relationship for WN (left) and
WC/WO (right) stars.  For WN stars $m = 0.54$ at log~$Z / Z_\odot > -2$ and
steepens to $m = 1.1$ at lower metallicity.  No models could be calculated for
log~$Z / Z_\odot < -3$.  For WC stars $m = 0.57$ for log~$Z / Z_\odot \geq -2$.
$\dot{M}$ then falls rapidly, finally flattening in the region log~$Z / Z_\odot
\leq -4$ to around $1.2\times10^{-8} \: M_\odot$ yr$^{-1}$ for $L_{th} = 0$ or
$6.5\times10^{-9} \: M_\odot$ yr$^{-1}$ for finite $L_{th}$.  The WC and WO
mass-loss rates are very similar.

\acknowledgements This work was performed 
under the auspices of the U. S. Department of Energy by the Los Alamos National
Security (LANS), LLC under contract No. DE-AC52-06NA25396.

\end{document}